\documentstyle[prc,aps,eqsecnum,epsfig]{revtex}            

\def\be{\begin{equation}}
\def\ee{\end{equation}}
\def\bea{\begin{eqnarray}}
\def\eea{\end{eqnarray}}
\def\D{\Delta}
\def\gm{\gamma_\mu}
\def\gn{\gamma_\nu}
\def\g5{\gamma_5}
\def\vep{\varepsilon}

\newcommand{\noi}{\noindent}

\def\Journal#1#2#3#4{{#1} {\bf #2} #3, (#4)}
\def\NPA{{ Nucl. Phys.} \bf A}
\def\NPB{{ Nucl. Phys.} \bf B}
\def\PRL{ Phys. Rev. Lett.\,\,}
\def\PRC{{ Phys. Rev.} \bf C}
\def\PRD{{ Phys. Rev.} \bf D}

\def\FBS{ Few--Body Systems\,\,}
\def\PLB{{ Phys. Lett.} \bf B}

\def\IJMPE{{ Int. J. Mod. Phys.} \bf E}

\def\HI{ Hyperfine Interactions\,\,}

\def\JPG{ J. Phys. \bf G}

\renewcommand{\thefootnote}{\fnsymbol{footnote}}

\begin{document}

\draft

\title{
{\large{\bf Gauge symmetric $\D$(1232) couplings and the radiative
muon capture in hydrogen}}}

\author{J.~Smejkal\footnote{Email address: smejkal@ujf.cas.cz}}
\address{Institute of Nuclear Physics, Academy of Sciences of the Czech Republic,
CZ--250 68 \v{R}e\v{z} n. Prague, Czechia }

\author{E.~Truhl\'{\i}k\footnote{Email address: truhlik@ujf.cas.cz}}
\address{Institute of Nuclear Physics, Academy of Sciences of the Czech Republic,
CZ--250 68 \v{R}e\v{z} n. Prague, Czechia
 \\
and
 \\
Institute for Nuclear Theory, University of Washington, Box
351550, Seattle WA, 98195, USA }


\maketitle

\begin{abstract}
\noi Using the difference between the gauge symmetric and standard
$\pi N \D$ couplings, a contact $\pi\pi NN$ term, quadratic in
the $\pi N \D$ coupling, is explicitly constructed. Besides, a
contribution from the $\D$ excitation mechanism to the photon
spectrum for the radiative muon capture in hydrogen is derived
from the gauge symmetric $\pi N \D$ and $\gamma N \D$ couplings.
It is shown for the photon spectrum, studied recently
experimentally, that the new spectrum is for the photon momentums
$k\,\ge\,60$ MeV by 4-10 \% smaller than the one obtained from
standardly used couplings with the on--shell $\D$'s.
\end{abstract}


\noi 


\renewcommand{\thefootnote}{\arabic{footnote}}
\setcounter{footnote}{0}

\section{Introduction}
\label{intro}

The photon spectrum in the radiative muon capture in hydrogen \be
\mu^-\, +\, p\, \longrightarrow \,\nu_\mu\, +\,\gamma\,+\, n\,,
\label{RMCp} \ee has recently been calculated by several authors
\cite{MMK,SADM,BHM,AMK,TK} in the search for a process enhancing
the high energy part of the photon spectrum, calculated earlier in
Ref.\,\cite{BF1}. It was concluded in \cite{BHM}, where the study
was made both within the small scale expansion scheme \cite{HHK}
and in the heavy baryon chiral perturbation theory \cite{BKM},
that a combination of various small effects could explain the
experimental spectrum \cite{TRIUMF1,TRIUMF2}. However, actual size
of some of these effects such as of the charge symmetry breaking,
are to be considered in a more detail. On the other hand, this
spectrum was calculated in \cite{TK} using amplitudes derived from
Lagrangians possessing the hidden local $SU(2)_L\times SU(2)_R$
symmetry \cite{STK,TSK}. In particular, the vertices containing
the $\D(1232)$ isobar field were chosen as \be {\cal
L}_{\,N\D\pi\rho a_{1}} \, = \, \frac{f_{\pi N \D}}
{m_{\pi}}\,\bar{\Psi}_{\mu}\vec{T}{\cal O}_{\mu\nu}(C(Z))\Psi\cdot
\left(\partial_{\nu}\vec{\pi} +2 f_{\pi}g_\rho\vec{a}_{\nu}\right)
-g_{\rho}\frac{G_{1}}{M}\bar{\Psi}_{\mu}\vec{T}{\cal
O}_{\mu\eta}(C(Y)) \gamma_{5}\gamma_{\nu}\Psi\cdot
\vec{\rho}_{\eta\nu}\,+\,h.\,c.   \label{LNDPRA1} \ee Here
$\vec{T}$ is the operator of the isospin $1/2\,\rightarrow\,3/2$
transition. The operator ${\cal O}_{\mu\nu}(C(B))$ is taken in a
form \cite{OO,DMW1,DMW2} \bea
{\cal O}_{\mu\nu}(C(B))\,&=&\,\delta_{\mu\nu}\,+\,C(B)\,
\gamma_\mu\,\gamma_\nu\,, \label{OmnB} \\
C(B)\,&=&\,-\left(\frac{1}{2}\,+\,B\right)\,.  \label{CeB} \eea
The parameters $Y$ and $Z$ do not influence the on--shell
properties of the $\D$ isobar, hence they are called off-shell
parameters. The vertices (\ref{LNDPRA1}) has frequently been used
\cite{OO,DMW1,DMW2,BDM,KA} to study the $\pi N$ reactions and the
pion photo- and electroproduction on nucleon and the parameters of
the model, including $Y$ and $Z$, were extracted from the data.

On the other hand, one can find also an attempt \cite{TE} to show
that the off--shell parameters are redundant within the framework
of effective fields theories. For this purpose, Tang and Ellis
consider the Lagrangian of the $\pi N \D$ system with the $\pi N
\D$ interaction of the type (\ref{LNDPRA1}). After integrating out
the $\D$ isobar field, they obtain a nonlocal $\pi N$ Lagrangian
where the $Z$ dependence is contained in couplings. This leads
them to conclude that these couplings can be redefined so that the
$Z$ dependence disappears and therefore, this parameter is
physically irrelevant. However, after finding that it is difficult
to manage the nonlocal part of the resulting Lagrangian, Tang and
Ellis return to the starting $\pi N \D$ Lagrangian containing the
$\D$ field explicitly and recommend to use it with some convenient
choice of the parameter $Z$, since it is not relevant to the
physics. On the other hand, they do not consider any  mechanism to
compensate the $Z$ dependence of the observables. Indeed, if such
an independence on a parameter should take place, one should
provide a mechanism to compensate it if it appears due to a
particular process, that can generally happen.

It can be seen \cite{TK,DMW2} that the $Z$ dependence of the
amplitudes appears in the form of contact terms. As it has
recently been discussed in Refs.\,\cite{PATI,PA}, the contact
nature of the $\D$ excitation amplitudes appears if the
interaction vertices contain the projection operators onto the
spin 1/2 space, that leads to the contribution of this space.
Indeed, the $\D$ propagator can be written in terms of projection
operators as \bea S^{\mu\nu}_F\,&=&\,\frac{1}{i\not
p+M_\D}[\delta_{\mu\nu}-\frac{1}{3}\gm\gn
+\frac{2}{3M^2_\D}p_\mu p_\nu+\frac{1}{3M_\D}(\gm p_\nu-\gn p_\mu)] \nonumber  \\
&=&\,-\frac{1}{i\not p+M_{\D}}(P^{3/2})_{\mu\nu}\,+\,
\frac{1}{\sqrt{3}M_{\D}}[(P^{1/2}_{12})_{\mu\nu}
\,+\,(P^{1/2}_{21})_{\mu\nu}] \,+\,\frac{2}{3M^2_{\D}}(i\not
p-M_{\D})(P^{1/2}_{22})_{\mu\nu}\,. \label{SMN} \eea In its turn,
the operator (\ref{OmnB}) reads \be {\cal
O}_{\mu\nu}(C(Z))\,=\,-(P^{3/2})_{\mu\nu}\,-\,(1+3C(Z))(P^{1/2}_{11})_{\mu\nu}
-(1+C(Z))(P^{1/2}_{22})_{\mu\nu}\,-\,\sqrt{3}C(Z)[(P^{1/2}_{12})_{\mu\nu}
\,+\,(P^{1/2}_{21})_{\mu\nu}]\,. \label{CZ} \\
\ee Explicit form of the projection operators can be found in
Ref.\,\cite{BDM}. It is seen that if the $\D$ propagator
$S^{\mu\nu}_F$ is sandwiched between the vertices of the type
(\ref{CZ}), the non--pole contribution is present. According to
\cite{PATI,PA}, this feature of the $\D$ interaction is related to
the change of the number of degrees of freedom in comparison with
the case allowed by the kinetic energy term of the $\D$
Lagrangian. As a remedy it is proposed in \cite{PATI,PA} that the
$\pi N\D$ and $\gamma N\D$ Lagrangians would possess the same
symmetry as the kinetic energy term of the $\D$ Lagrangian that is
invariant under a kind of the gauge transformation \be
\Psi_\mu(x)\,\rightarrow\,\Psi_\mu(x)\,+\,\partial_\mu\,\xi(x),
\label{GT} \ee where $\Psi_\mu(x)$ is the $\D$ isobar field and
$\xi(x)$ is a spinor.

In Refs.\,\cite{PATI,PA}, new $\pi N \D$ and $\gamma N \D$
Lagrangians are proposed. They possess the property that with the
proper choice of couplings, the new and traditional Lagrangians
provide identical $\pi N \D$ and $\gamma N \D$ vertices for the
on--shell particles. Further, using the redefinition (\ref{GT}),
it is shown in \cite{PA} that the new and traditional $\pi N \D$
couplings differ by a contact term quadratic in the coupling
constant that can be associated with the contribution of the 1/2
spin space involved due to the traditional $\pi N \D$ coupling
\footnote[1]{Recent review of some aspects of the theory of massive
Rarita-Schwinger fields can be found in \cite{TP}.}.

In Sect.\,\ref{CH1}, we show how one can construct such a contact
term directly. For this purpose, we first use an identity to show
that the new and traditional $\pi N \D$ couplings differ by a sum
of the $\pi N \D$ terms that vanish for the $\D$ isobar on--shell.
Next we construct, in the tree approximation, the contribution of
the $\D$ excitation to the $\pi N$ scattering amplitude and we
show that these $\pi N \D$ terms give rise to a contact term
quadratic in the $\pi N \D$ coupling constant. in
Sect.\,\ref{CH2}, we use the new $\pi N \D$ and $\gamma N \D$
couplings to calculate the $\D$ excitation contribution to the
photon spectrum for the reaction (\ref{RMCp}). We show for the
recently measured spectrum \cite{TRIUMF1,TRIUMF2} that it is
suppressed in comparison with the one obtained earlier with the
use of the traditional couplings. In Sect.\,\ref{CH3}, we discuss
the obtained results and we present our conclusions.

\section{The $\pi N \D$ Couplings}
\label{CH1}

According to Ref.\,\cite{PATI}, the gauge symmetric $\pi N \D$
coupling is \be {\cal L}^{g.s.}_{\,\pi N\D} \, = \,
f\,\vep_{\mu\nu\alpha\beta}
\,(\partial_\mu\bar{\Psi}_{\nu})\,\vec{T}\g5\gamma_\alpha\,\Psi\cdot
(\partial_{\beta}\vec{\pi})\,+\,h.c.\,.  \label{LGS} \ee With the
choice \be f\,=\,\frac{f_{\pi N \D}}{m_\pi M_\D}\,,  \label{FGS}
\ee and for the $\D$ isobar on--shell ($Z=-1/2$), this coupling is
equivalent to the traditional one \be {\cal L}_{\,\pi N\D}(Z=-1/2)
\, = \,\frac{f_{\pi N \D}} {m_{\pi}}\,\bar{\Psi}_{\mu}\vec{T}{\cal
O}_{\mu\nu}(C(Z=-1/2))\Psi\cdot (\partial_{\nu}\vec{\pi})\,+\,h.c.\,.
\label{LTR} \ee Using the identity \be
\vep_{\mu\nu\alpha\beta}\g5\gamma_\alpha\,=\,-\delta_{\mu\nu}\gamma_\beta\,+\,
\delta_{\beta\nu}\gm\,-\,\delta_{\beta\mu}\gn\,
+\,\gn\gm\gamma_\beta\,,  \label{IDE} \ee one can show that \be
{\cal L}^{g.s.}_{\,\pi N\D} \, = \,{\cal L}_{\,\pi N\D}(Z)\,
+\,\delta\,{\cal L}_{\,\pi N\D}(Z)\,,  \label{LGSLTR} \ee where
\bea \delta\,{\cal L}_{\,\pi
N\D}(Z)\,&=&\,f\left\{\,-(\partial_\mu\,{\bar\Psi}_\mu)\vec T
\gn\,\Psi\,-\,(\partial_\nu\,{\bar\Psi}_\alpha\,\gamma_\alpha)\vec
T\,\Psi \,+\,(\partial_\mu\,{\bar\Psi}_\alpha)\,\gamma_\alpha{\vec
T}\gm\gn\,\Psi \right.
\nonumber \\
&& \left.
\,+\,{\bar\Psi}_\nu[(\gm\stackrel{\leftarrow}{\partial}_\mu)\,-\,
M_\D]\,{\vec T}\, \Psi\,-\,C(Z)\,M_\D\,{\bar\Psi}_\mu \gm\,{\vec
T}\gn\,\Psi\,\right\} \,\cdot\,\partial_\nu\vec \pi\,.
\label{DLPND} \eea It holds for the $\D$ isobar on--shell that \be
\partial_\mu\Psi_\mu(x)\,=\,\gm\Psi_\mu(x)\,=\,[\not\partial\,+\,M_\D]\Psi(x)\,=\,0\,.
\label{TEQ} \ee It follows from these equations that for the $\D$
isobar on--shell $\delta\,{\cal L}_{\,\pi N\D}(Z)=0$.

\begin{figure}[h!]
\centerline{ \epsfig{file=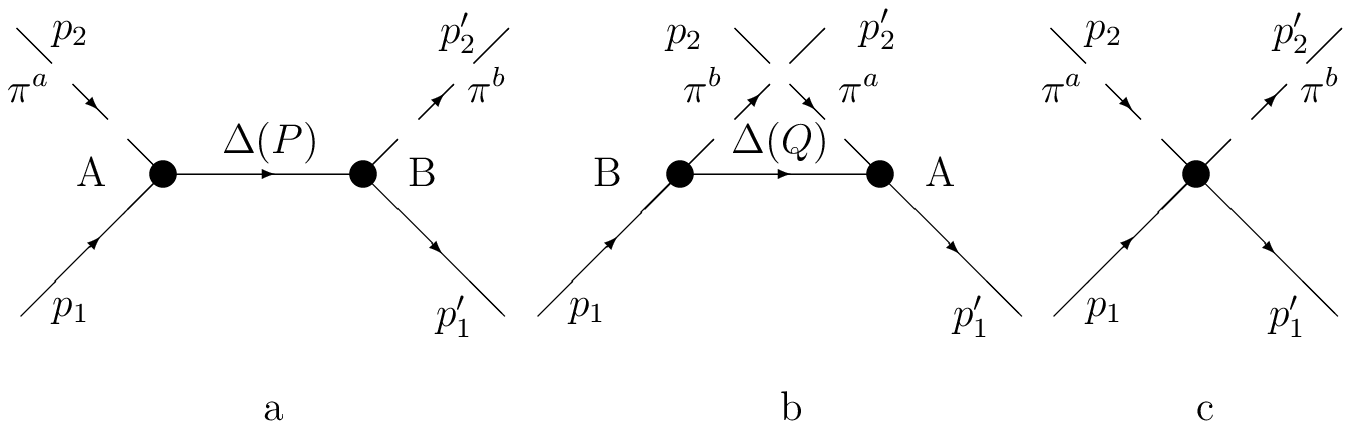} }
\vskip 0.4cm \caption{The $\pi N$ scattering amplitudes in the
tree approximation; a, b- the $\D$ excitation amplitudes, c- the
contact term. } \label{figg1}
\end{figure}

In the tree approximation, the $\pi N$ scattering via the $\D$
isobar excitation is described by the Feynman graphs of
Figs.\,\ref{figg1}a and \ref{figg1}b.
One can calculate the S-matrix element corresponding to
Fig.\,\ref{figg1}a either using the left hand side of
Eq.\,(\ref{LGSLTR}) or, equivalently, its right hand side. If one
considers a part of the S-matrix element, $S_p$, given by the sum
of the partial S-matrix elements, calculated with the choice \bea
A\,&=&\,{\cal L}_{\,\pi N \D}(Z),\hskip 2.5mm\qquad
B\,=\,\delta\,{\cal L}^+_{\,\pi N\D}(Z),
\\ \label{LDLP}
A\,&=&\,\delta\,{\cal L}_{\,\pi N\D}(Z),\qquad B\,=\,{\cal
L}^+_{\,\pi N \D}(Z),
\\ \label{DLLP}
A\,&=&\,\delta\,{\cal L}_{\,\pi N\D}(Z),\qquad B\,=\,\delta\,{\cal
L}^+_{\,\pi N \D}(Z), \label{DLDLP} \eea one obtains the
difference between the S-matrix elements, calculated first with
the new Lagrangian ${\cal L}^{g.s.}_{\,\pi N\D}$ and then only
with the traditional Lagrangian ${\cal L}_{\,\pi N\D}(Z)$.
Explicit calculations yield $S_p$ in the form of the $\pi\pi NN$
contact  graph of Fig.\,\ref{figg1}c. Defining \be
S_p\,=\,-i(2\pi)^4\,\delta^{(4)}(P_f\,-\,P_i)(\chi^b)^+\,T^{ba}_p(s)\,\chi^a\,,
\label{Sp} \ee
we obtain for the amplitude $T^{ba}_p(s)$ the
following equation
\bea T^{ba}_p(s)\,&=&\,f^2\,M_\D\,{\bar
u}(p'_1)\big<\,p'_{2\nu}[-\delta_{\nu\mu}+\frac{1}{3}
\gn\gm+\frac{i}{3M_\D}(3{\not P}\delta_{\nu\mu}+\gn{\not
P}\gm-P_\nu\gm-P_\mu\gn)]
p_{2\mu} \big. \nonumber \\
&&\big.\,+\frac{2}{3} C(Z)\,\left\{\,{\not p}'_2{\not p}_2
+\frac{i}{M_\D}[(p'_2\cdot P){\not p}_2+{\not p}'_2(P\cdot
p_2)]\,\right\}
\big.\nonumber \\
&&\big.\,+\frac{2}{3} C^2(Z)\,{\not p}'_2(2+i\frac{\not P}{M_\D})
{\not p}_2\,\big>\,\big(T^+\big)^b\,T^a\,u(p_1)\,.
\label{TBAP1} \eea
Here $P=p_1+p_2=p'_1+p'_2$,
$\big(T^+\big)^b\,T^a=\frac{2}{3}\delta_{ba}-
\frac{1}{3}i\vep^{bac}\tau^c$ and $\tau^c$ are the isospin Pauli
matrices. In deriving Eq.\,(\ref{TBAP1}) we supposed $C(Z)$ to be
a real function of the variable $Z$.

The amplitude $T^{ba}_p(s)$ corresponds to an effective contact
Lagrangian
\bea {\cal L}_{\pi\pi
NN}(Z)\,&=&\,-f^2\,M_\D\,(\partial_\nu \pi^b)\,\bar{\Psi}\left\{
\delta_{\nu\mu}\,-\,\frac{1}{3}\left[\,1+2C(Z)+4C^2(Z)\,\right]\,\gn\gm
\,-\,\frac{1}{M_\D}\delta_{\nu\mu}\not\partial \right. \nonumber \\
&&\left.\,-\,\frac{1}{3M_\D}\left[\,
1+2C^2(Z)\right]\gn\gm\not\partial\,+\,\frac{1}{3M_\D}\left[\,1-2C(Z)\right]
(\gn\partial_\mu\,+\,\gm\partial_\nu)
\right\}\,\big(T^+\big)^b\,T^a\, \left[\Psi(\partial_\mu
\pi^a\right]\,.  \label{LCON}
\eea
Let us write this equation in the form
\be
{\cal L}_{\pi\pi NN}(Z)\,\equiv\,-f^2\, M_\D\,(\partial_\nu \pi^b)\,\bar{\Psi}\,
{\cal T}_{\nu\mu}(C(Z))\,\big(T^+\big)^b\,T^a\, \left[\Psi(\partial_\mu
\pi^a\right]\,.  \label{LCON1}
\ee
The straightforward calculations  yield the amplitude ${\cal T}_{\nu\mu}(C(Z))$
in the factorized form
\be
{\cal T}_{\nu\mu}(C(Z))\,=\,{\cal O}_{\nu\alpha}(C(Z))\,
{\cal T}_{\alpha\beta}(C(Z)=0)\,{\cal O}_{\beta\mu}(C(Z))\,.  \label{FACT}
\ee
It follows from the form of the operator vertex (\ref{OmnB}) that
\be
{\cal O}_{\nu\alpha}(a)\,{\cal O}_{\alpha\mu}(b)\,=\,
{\cal O}_{\nu\alpha}(b)\,{\cal O}_{\alpha\mu}(a)\,=\,
{\cal O}_{\nu\mu}(a+b+4ab)\,=\,\delta_{\nu\mu}\,,  \label{Oab}
\ee
if the parameters $a$ and $b$ satisfy the equation
\be
a+b+4ab\,=\,0\,.  \label{eab}
\ee
It can be seen from Eqs.\,(\ref{Oab}) and (\ref{eab}) that the amplitude
${\cal T}_{\nu\mu}(C(Z))$ can be factorized in an infinite number of forms.
Choosing $a=-1$ and $b=-1/3$, we can write Eq.\,(\ref{FACT}) as
\bea
{\cal T}_{\nu\mu}(C(Z))\,&=&\,{\cal O}_{\nu\alpha}(C(Z))\,[\delta_{\alpha\beta}
{\cal T}_{\beta\lambda}(C(Z)=0)\delta_{\lambda\tau}]\,{\cal O}_{\tau\mu}(C(Z)) \nonumber \\
&=&\,{\cal O}_{\nu\alpha}(C(Z))\,[({\cal O}_{\alpha\eta}(-1/3)
{\cal O}_{\eta\beta}(-1))\,{\cal T}_{\beta\lambda}(C(Z)=0)\,({\cal O}_{\lambda\sigma}(-1)
{\cal O}_{\sigma\tau}(-1/3))]\,{\cal O}_{\tau\mu}(C(Z)) \nonumber \\
&=&\,{\cal O}_{\nu\eta}(-\frac{1}{3}(1+C(Z)))\,[{\cal T}_{\eta\sigma}(C(Z)=-1)]\,
{\cal O}_{\sigma\mu}(-\frac{1}{3}(1+C(Z)))
\,.  \label{FACTP}
\eea
In Ref.\,\cite{PA}, applying the field
redefinition (\ref{GT}) with a particular choice of the field
\be
\partial_\mu\,\xi\,=\,-\frac{1}{M_\D}{\cal O}_{\mu\rho}(C(Z)=-1/3){\cal O}_{\rho\nu}(C(Z))
T^a\Psi\partial_\nu\phi^a\,,  \label{XIP}
\ee
in the $Z$ dependent Lagrangian (\ref{LTR}), besides the gauge symmetric Lagrangian (\ref{LGS}),
an effective contact Lagrangian ${\cal L}_{\pi\pi NN}(Z)$
of the form (\ref{LCON1}) was obtained, where the
amplitude ${\cal T}_{\nu\mu}(C(Z))$ is given by Eq.\,(\ref{FACTP}).

In the next section, we apply the gauge symmetric Lagrangians to the calculation
of the contribution of the $\Delta$ excitation process to the photon spectrum in
the radiative muon capture in the hydrogen.

\section{The photon spectrum in the radiative muon capture in hydrogen}
\label{CH2}

The new Lagrangians, needed for the calculations of the photon
spectrum, that are derived from the gauge symmetric ones
\cite{PATI} read \be {\cal L}^{g.s.}_{\,\pi N\D a_1} \, =
\,f\,\vep_{\mu\nu\alpha\beta}
\,[(\partial_\mu\bar{\Psi}_{\nu})\,\vec{T}\g5\gamma_\alpha\,\Psi]\cdot
(\partial_{\beta}\vec{\pi}\,+\,2f_\pi g_\rho {\vec
a}_\beta)\,+\,h.c.\,,   \label{LGSPND} \ee \be {\cal
L}^{g.s.}_{\,\rho N\D } \, =
\,f_P\,g_\rho\,\vep_{\mu\nu\alpha\beta}
\,[(\partial_\mu\bar{\Psi}_{\nu})\,\vec{T}\gamma_\alpha\gamma_\lambda\,\Psi]\cdot
{\vec \rho}_{\lambda
\beta}\,+\,f_P\,g_\rho\,[(\partial_\mu\bar{\Psi}_{\nu}
-\partial_\nu\bar{\Psi}_{\mu}){\vec T}\g5\gm\gamma_\lambda\,\Psi]
\cdot{\vec \rho}_{\lambda\nu}\,+\,h.c.\,.  \label{LGSRND} \ee Here
$f_P=\frac{G_1}{MM_\D}$ is obtained from the condition that the
new (\ref{LGSRND}) and the standard $\rho N\D$ couplings
(\ref{LNDPRA1}) are equivalent for the $\D$ isobar on--shell. The
notations of this section coincide with the notations of
Refs.\,\cite{TK,STK}, where one can also find more detailed discussion
of the reaction (\ref{RMCp}).

\begin{figure}[h!]
\centerline{ \epsfig{file=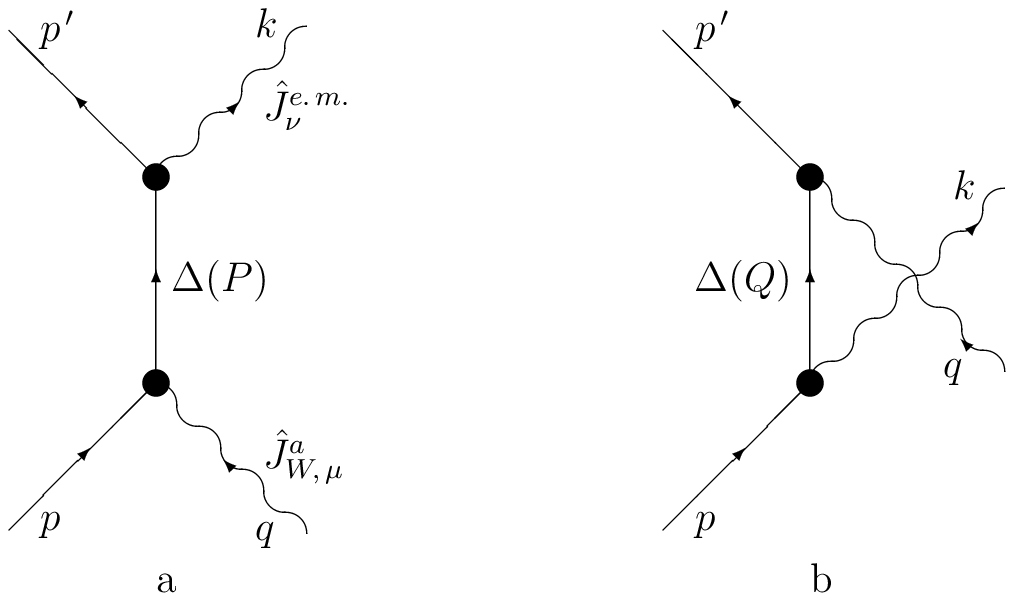} }
\vskip 0.4cm \caption{The $\D$ excitation amplitudes contributing
to the radiative muon capture in hydrogen. The weak hadron current
$\hat{J}^{a}_{W,\,\mu}$, interacting with the nucleon, is exciting
it and the $\D$ isobar in the intermediate state appears that
decays into the final state nucleon and photon. } \label{figg2}
\end{figure}

The contribution from the $\D$ excitation processes to the photon
spectrum for the reaction (\ref{RMCp}) is presented in
Fig.\,\ref{figg2}.

From various form factors, calculated in Sect.\,II.B of
Ref.\,\cite{TK}, we need consider \be \D
g_2\,=\,-\frac{8}{9M_\D}\,f_{\pi N\D}\,G_1\,\frac{f_\pi}{m_\pi}
\,\eta\,k\,\left[-(1+2R)+2(1-2R)C(Y)+2(1-R)C(Z)
+4(2-R)C(Y)C(Z)\right] \,,  \label{DG2G} \ee and \be \D
g_3\,=\,-\frac{16}{9}\lambda\,f_{\pi
N\D}\,G_1\,\frac{f_\pi}{m_\pi}
\,\eta\,k\,\frac{1}{M_\D-M}\left\{1+(1-R)\left[C(Y)+C(Z)+2(2+R)C(Y)C(Z)
\right]\right\}\,,   \label{DG3G} \ee Here $\lambda$ is the photon
polarization, $k$  is the photon momentum, $\eta=\frac{m_\mu}{2M}$
and $R=M/M_\D$. As it can be seen from Eq.\,(\ref{DG2G}), this
form factor is of the contact origin, whereas inspecting of the
Eq.\,(\ref{DG3G}) shows that the dependence of the form factor on
the off--shell parameters $Y$ and $Z$ is entirely located in the
contact part of the form factor.

Using new Lagrangians (\ref{LGSPND}) and (\ref{LGSRND})  and
performing calculations, identical to those presented in
Sect.\,II.B of \mbox{Ref.\,\cite{TK}}, one obtains \be \D
g_3^{g.s.}\,=\,-\frac{16}{9}\lambda\, f_{\pi
N\D}\,G_1\,\frac{f_\pi}{m_\pi}
\bigg(\frac{M}{M_\D}\bigg)^2\,\eta\,k\,\frac{1}{M_\D-M}\,.
\label{DG3} \ee

In contrast to the calculations with the standard couplings, now
the form factor $\D g^{g.s.}_3$ contains only the $\D$ pole contribution.
This follows from the fact that if the propagator (\ref{SMN}),
given in terms of the projection operators, is sandwiched
between gauge symmetric Lagrangians, only the part proportional
to the projection operator $(P^{3/2})_{\mu\nu}$, accompanied
by the $\D$ isobar pole, contributes to the matrix element \cite{PA}.
This simplifies the calculations of matrix elements considerably.

Let us write down the contribution $\D g^0_3$, given in
Eq.\,(\ref{DG3G}), for the $\D$ isobar on--shell ($Y=Z=-1/2$) \be
\D g^0_3\,=\,-\frac{16}{9}\lambda\, f_{\pi
N\D}\,G_1\,\frac{f_\pi}{m_\pi} \,\eta\,k\,\frac{1}{M_\D-M}\,,
\label{DG30} \ee and calculate the difference with the form factor
$\D g^{g.s.}_3$ from Eq.\,(\ref{DG3}) \be \D g^{g.s.}_3\,-\,\D
g^0_3\,=\,-\frac{16}{9}\lambda\, f_{\pi
N\D}\,G_1\,\frac{f_\pi}{m_\pi}
\,\eta\,\frac{k}{M_\D}\,\bigg(1\,+\,\frac{M}{M_\D}\bigg)\,.
\label{DG3MG30} \ee As it is seen, the difference is a contact
term, in agreement with the more general discussion that took
place above.

\begin{figure}[h!]
\centerline{ \epsfig{file=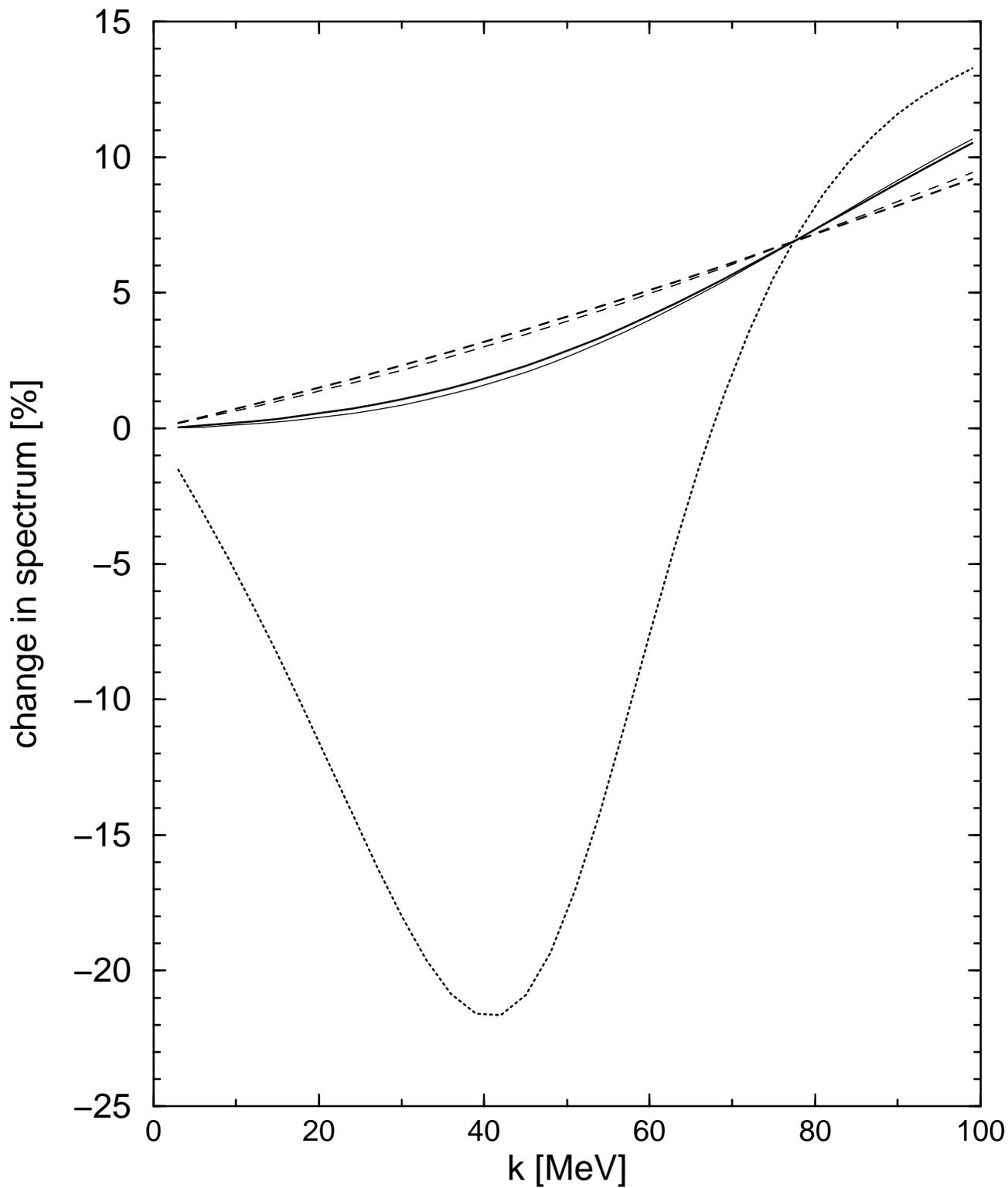} }
\vskip 0.4cm \caption{The change in the photon spectrum. Solid
curve- the spectrum measured in the TRIUMF experiment; dashed
curve- the spectrum for the muon--hydrogen triplet state; dotted
curve- the spectrum for the muon--hydrogen singlet state.}
\label{figg3}
\end{figure}

In Fig.\,\ref{figg3}, we present the change in the photon spectrum calculated as\\
\begin{center}
{[spectrum calculated using $\D g^0_3$ from
Eq.\,(\ref{DG30})]-[spectrum calculated using $\D g^{g.s.}_3$ from
Eq.\,(\ref{DG3})]}/ [spectrum calculated using $\D g^0_3$ from
Eq.\,(\ref{DG30})].
\end{center}
Other contributions are the same as in Ref.\,\cite{TK}. The
spectrum measured in the TRIUMF experiment \cite{TRIUMF1,TRIUMF2}
is given as \be S_T\,=\,0.061 S_s\,+\,0.854 S_o\,+\,0.085 S_p\,.
\label{TS} \ee Here $S_s$, $S_o$ and $S_p$ correspond to the
muon-hydrogen singlet system, and to the ortho- and paramolecular
$p\mu p$ states, respectively. As it is seen from
Fig.\,\ref{figg3}, the spectrum $S_T$, calculated with  $\D g^{g.s.}_3$
from Eq.\,(\ref{DG3}) is in the region \mbox{$k\,>\,60$ MeV}
suppressed, in comparison with the spectrum obtained using the
traditional couplings. It means that the new couplings cannot
resolve the "$g_P$ puzzle" either. Minor difference between the
curves of the same sort comes from omitting the form factor $\D
g_2$ of Eq.\,(\ref{DG2G}) in the calculations.

It is seen from Eqs.\,(\ref{DG3}) and (\ref{DG30}) that the form
factors $\D g^{g.s.}_3$ and $\D g^0_3$ differ by the factor
$(M/M_\D)^2\approx 0.58$. Such a factor will appear also in the
meson exchange current operators with the $\D$ excitation. On the
other hand, a suppression factor 0.8 has been found to be needed
\cite{Sch14} to reduce the effect of the weak axial meson exchange
currents with the $\D$ excitation in order to explain the
experimental value of the Gamow--Teller matrix element for the
triton $\beta$ decay, if the value of the constant $f_{\pi N\D}$
is taken from the constituent quark model. In other words it means
that effectively the value of the constant $f_{\pi N\D}$ turns out
to be unrealistically small or one should speculate about other
processes effectively suppressing the meson exchange current
effect \cite{MSRKV}. If these weak axial exchange currents are
constructed from the new Lagrangians, the factor $(M/M_\D)^2$
appears naturally and the value of the constant $f_{\pi N\D}$ can
be taken larger and therefore, more realistic. Simultaneously,
such a factor will appear also in the vector meson exchange
currents with the $\D$ excitation. However, the precise data on
the radiative capture of neutrons by protons do not demand any
damping of the vector meson exchange currents effect \cite{NSGK}
and the capture rate for the reaction $\mu^- + ^{3}He
\rightarrow\nu_\mu + ^{3}H$, that has been measured in the precise
experiment \cite{VOR,ACK}, is by the suppressed weak axial
exchange currents underestimated \cite{MSRKV}. Precise data,
expected from the experiments on the ordinary muon capture in
hydrogen and deuterium \cite{MUCAP}, will be for the axial sector
of the weak nuclear interaction very helpful.

\section{Results and conclusion}
\label{CH3}

Here we study some aspects of new  $\pi N \D$ and $\gamma N \D$
couplings  that have recently been proposed \cite{PATI} by
Pascalutsa and Timmermans. In comparison with the traditional
couplings, the new ones possess an additional gauge symmetry
(\ref{GT}) that is present in the kinetic energy term of the $\D$
Lagrangian. This symmetry guarantees that the couplings have the
same $\D$ degrees of freedom as the kinetic energy term. As a
consequence, the amplitudes of the processes with the $\D$
excitation in the intermediate state do not contain the contact
terms coming from the 1/2 spin space.

In Sect.\,\ref{CH1}, we study the difference between the
traditional  and gauge symmetric $\pi N \D$ couplings. Using an
algebraic identity between the gamma matrices, we first express
the new coupling as the sum of the traditional coupling and of
terms that are zero for the $\D$ isobar on--shell. The $\pi N$
scattering amplitude, constructed from these terms, is a contact
term, quadratic in the coupling constant. In Ref.\,\cite{PA}, such
a term was obtained by imposing the symmetry condition (\ref{GT})
on the traditional coupling. We also show that this contact term
can be factorized in an infinite number of forms.

In Sect.\,\ref{CH2}, we employ the gauge symmetric $\pi N \D$ and
$\gamma N \D$ couplings \cite{PATI} to calculate the photon
spectra for the radiative muon capture in hydrogen. As a result,
the new form factor $\D g_3$ contains only the $\D$ isobar pole
contribution. This form factor differs from the old one,
calculated for the $\D$ isobar on--shell, by the damping factor
$(M/M_\D)^2\approx 0.58$. Consequently, the new photon spectrum,
corresponding to the spectrum measured in the TRIUMF experiment,
is suppressed in the region $k\,>\,60$ MeV, in comparison with the
photon spectrum, calculated from the traditional couplings.
Therefore, the problem of extraction of the induced pseudoscalar
form factor $g_P$ from the photon spectrum in the radiative muon
capture in hydrogen cannot be solved by employing the gauge
symmetric $\pi N \D$ and $\gamma N \D$ couplings.

Let us note that the damping factor $(M/M_\D)^2$ will be also
present in the meson exchange current operators with the $\D$
isobar excitation, if for the construction the gauge symmetric
couplings are used. However, the comparison of the existing
calculations with the present data on the weak and electromagnetic
reactions in few--nucleon systems does not allow to decide
uniquely, if this factor is needed or not.

\section*{Acknowledgments}

This work is partially supported by the grant GA \v{C}R
202/03/0210 and by the ASCR project K1048102. One of us (E.T.)
thanks  the Institute for Nuclear Theory at the University of
Washington for its hospitality and the Department of Energy for
partial support during the initial stage of this work.

\end{document}